# Extreme diffusion limited electropolishing of niobium radiofrequency cavities

Anthony C. Crawford

Fermilab, Box 500, MS316, Batavia, IL 60510, USA

**Abstract**

A deeply modulated, regular, continuous, oscillating current waveform is reliably and repeatably achieved during electropolishing of niobium single-cell elliptical radiofrequency cavities. Details of the technique and cavity test results are reported here. The method is applicable for cavity frequencies in the range 500 MHz to 3.9 GHz and can be extended to multicell structures.

*Keywords*: SRF; niobium; cavity; electropolish; residual resistance

## 1. Introduction

Intermittent, diffusion-limited electropolishing of niobium for superconducting radiofrequency (SRF) cavities was first described by researchers at the Siemens Corporation [1]. Continuous electropolishing of rotating niobium cavities, using principles developed at Siemens, was introduced by researchers at KEK and Nomura Plating [2]. Reference [2] is a particularly valuable source for the history of the development of SRF cavity electropolishing.

An important aspect of the Siemens research is that best results for electropolishing of niobium are obtained by operating in a state of large current oscillations. A graph of current verses time is reprinted from [1] in Fig. 1. A brief explanation of the cause of the current instability is as follows: The electrolyte mixture contains sulfuric acid ($H_2SO_4$) and hydrogen fluoride (HF). Sulfuric acid acts as an oxidizing agent, forming niobium pentoxide ($Nb_2O_5$), an electrical insulator. HF dissociates niobium pentoxide. Under the influence of the electrical potential, both processes occur simultaneously, and at a reasonably high rate. As electropolishing proceeds, a dielectric layer of niobium salts develops at the niobium (anode) surface. It takes time for HF from the electrolyte to diffuse through the dielectric layer and produce a concentration high enough to break down the $Nb_2O_5$ layer, thus allowing current to reach a maximum. The HF concentration is then locally depleted, and current is minimized until

A technique for achieving a regular, periodic, deeply modulated current waveform during electropolishing of fully assembled cavities has been developed at Fermilab and will be referred to in this report as the "cold EP" method. The buildup of a high quality dielectric layer and control of the rate of diffusion of HF through this layer are of prime importance in achieving the goal. The starting point for the work is an electropolishing tool similar to the KEK horizontal electropolishing system as per K. Saito [2]. In the next section the critical process parameters will be identified and specified.

## 2. Electropolishing technique

### 2.1. Electrolyte

The electrolyte used at Fermilab is a 13.5 to 1 (by volume) mixture of concentrated sulfuric acid ($H_2SO_4 > 96\%$ by weight) and hydrofluoric acid (HF~ 70% by weight). The resulting mixture differs from that used in most niobium cavity electropolish operations in that the concentration of water is lower. The ratio of HF to $H_2SO_4$ is similar to traditional mixtures. Water interferes with the formation of the dielectric layer at the niobium surface. The lower the concentration of water, the better will be the quality of the insulating layer. As water concentration increases, current increases and the depth of the current oscillations decreases, until the dielectric layer and diffusion limited operation are lost.

Hydrofluoric acid with 70% HF concentration is not present, nor is it handled at Fermilab. Fermilab purchases pre-mixed electrolyte in containers of 120 to 200 liters from Acid Products Corporation of Chicago, Illinois. Significantly less heat is generated when mixing 70% hydrofluoric acid with sulfuric acid than with 48% hydrofluoric acid because of the lower water content. Lower heat generation during mixing results in improved thermodynamic control of the mixing process and aids in minimizing HF lost due to evaporation.

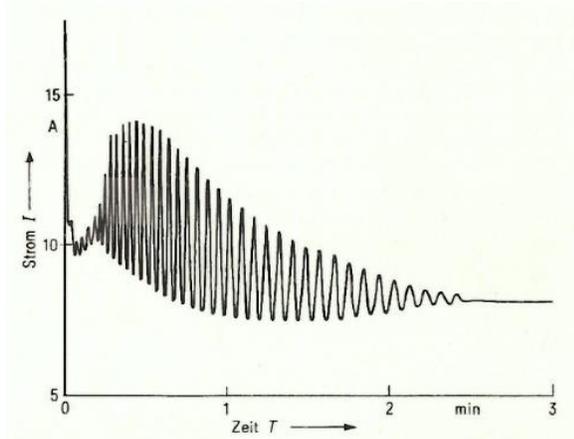

Fig. 1. Current oscillations. The horizontal axis is time and the vertical axis is current.

enough time has passed to allow diffusion to build up the HF concentration at the niobium pentoxide surface. The electrochemical cycle then repeats indefinitely, provided the bulk electrolyte solution can supply the required amount of HF.

The presence of the dielectric layer has the additional beneficial effect that a significantly large amount of the voltage drop from the cathode to the anode surface is across the layer. This is what makes it possible to have a reasonably large amount of material removal at the large diameter equators of elliptical cavities at the same time as at the smaller diameter irises without the necessity of a conformal cathode that maintains a uniform spacing to the anode.

Since the introduction of the current oscillation concept, it has been a goal to achieve continuous large amplitude oscillations for elliptical cavity electropolish procedures. The best results observed to date in this regard have been with polishing of open half cells [3]. This was due in large part to the use of a conformal cathode and because of the upward facing orientation of the niobium surface. A high density dielectric layer forms easily on such a surface due to the effect of gravity. A downward facing surface will not build up a layer that provides an optimal voltage drop with the type of electrolyte mixture used here.

## 2.2. Fluid dynamics of the electrolyte

The geometry of the single-cell electropolish assembly is shown in Fig 2. The cavity is axisymmetric. The cathode is made of aluminum with a minimum of 99.5% purity. The goal for electrolyte flow is to provide adequate HF concentration at the interface between the electrolyte and the dielectric layer, while causing minimal disturbance to the layer due to turbulence. Electrolyte introduction holes are located only on the top half of the hollow cylindrical cathode. This is so that flow is not directed towards the niobium surface. Multiple holes are used in the cathode in order to minimize electrolyte injection velocity while maintaining cathode structural strength and rigidity. The flow of electrolyte is adjusted to approximately one liter per minute for a 1.3 GHz 1-cell cavity.

Fig. 3 shows a photograph of the central section of the cathode. The masking material is polytetrafluoroethylene (PTFE) tape. The need for the use of masking is questionable, especially at low cathode temperature. It is likely that the results reported here would be obtained if no masking were in use. (The reader should note that for the case of cavities equipped with high order mode couplers, masking of the cathode at the coupler location is beneficial due to the very small gap between the coupler and the cathode. Electrical breakdown of the electrolyte can occur, especially at high electrolyte temperature.) No screen is used to contain hydrogen gas bubbles produced at the cathode. Hydrogen gas is continuously flushed from the cavity with a one liter per minute flow of nitrogen.

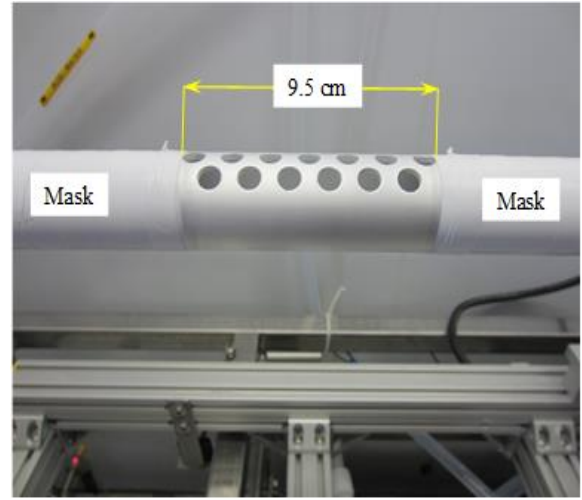

Fig. 3. The cathode, showing holes for electrolyte flow.

The level of electrolyte is maintained at approximately 60% of the volume of the cavity. This provides enough liquid inside the cavity to cover the top of the cathode with approximately 5mm of electrolyte. The cavity rotates slowly about its axis at an angular velocity of one revolution per minute for a 1.3 GHz cavity. This is equivalent to a surface speed at the equator of 1.1 cm/sec.

## 2.3. Temperature control

The electrolyte serves two functions: (1) to provide a fresh supply of HF and $H_2SO_4$ for the electrochemical reaction, and (2) as a coolant to transport heat away from the niobium surface. In the first generation of cavity electropolish tools, the electrolyte was responsible for all of the heat removal. For the system described in this report, the functions have been separated to a large extent, allowing for minimal electrolyte flow while maintaining the cavity at low temperature. In the Fermilab system the cavity exterior surface is divided into three zones where the temperature of the exterior cavity wall is controlled by application of cold water flow. This arrangement is shown in Fig. 4 and Fig. 5. The zone separators are flexible plastic disks formed from polycarbonate sheet. They prevent cooling water from moving from one zone to another.

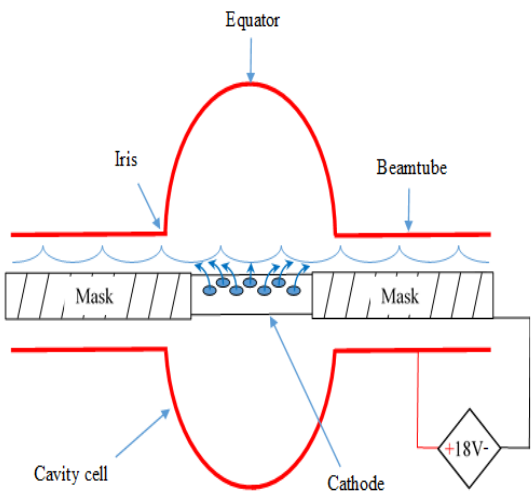

Fig. 2. Elliptical 1-cell cavity cathode arrangement.

The electrolyte is chilled to 8 C and is continuously introduced through the cathode holes at this temperature. One half of the electrolyte flow exits the cavity through each beamtube. By adjusting the flow rate of the 5 C external cavity cooling water independently for each zone, the temperature of the equator is maintained at 16 C while the beamtubes are at 6 C. The temperature of the electrolyte is 10 C as it exits the cavity. The typical average power generated at the niobium surface is approximately 270 watts. Approximately 2 liters of cooling water are used per minute for each beamtube and 250 milliliters per minute on the cavity cell. The dissipated electrical power, and therefore, the amount of water required, especially for the cavity cell, will vary significantly depending on the concentration of HF in the electrolyte. Temperatures reported here are measured by bare thermocouple junctions in good thermal contact with the exterior surface of the cavity. The thermocouple junctions are thermally insulated from the flow of water and their readings are a reasonable representation of the temperature of the bulk niobium. Because the thermocouples necessarily rotate with the cavity, their signals are read by means of radio transmission.

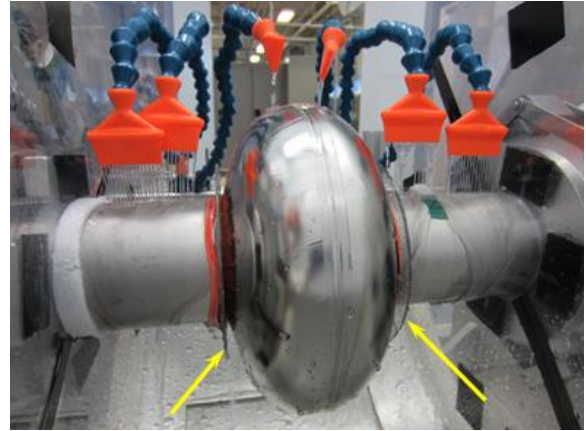

Fig. 5. Elliptical 1-cell external water cooling photograph. The zone separators are the transparent plastic disks at the iris locations. Yellow colored arrows indicate their positions.

*2.4. Voltage*

As stated in Diepers, et al., "optimum polishing conditions are not governed by the plateau voltage in the current-voltage characteristic." Optimal polishing is characterized by regular, periodic current oscillations, and these can be obtained within a wide range of values for voltage. For 1.3 GHz TESLA shaped cavities, Fermilab uses the value of 18 V between the cathode tube and the cavity (anode). This value was chosen because of its favorable "throwing power," i.e., the ability to do a good job of polishing anode surfaces both near and far from the cathode.

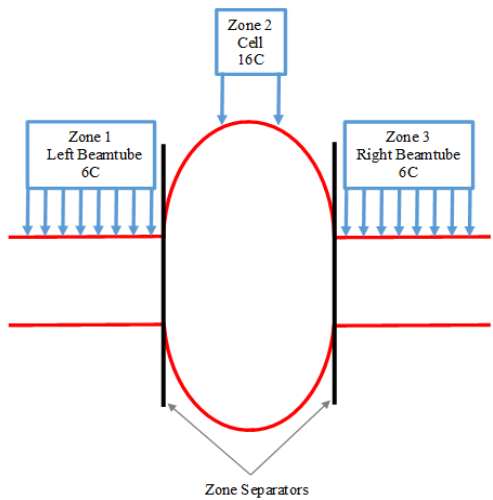

Fig. 4. Elliptical 1-cell external water cooling diagram.

Increasing the voltage, up to 18 V, tends to minimize the difference between material removal at the iris and the equator. At 18 V, macroscopic "levelling" is good, and there is no evidence of microscopic etching over the interior surface of the cavity cell, or in the beamtubes.

*2.5. Post electropolish HF rinse*

It has been shown that electropolished cavities can have niobium oxide crystals adhering to the interior surface [4]. In order to remove as many sites like this as possible, Fermilab has adopted the practice of allowing the electrolyte to circulate through the rotating cavity for 30 minutes after the polishing procedure has concluded. The HF content in the electrolyte is high enough to dissociate the oxide, and in principle, remove the crystalline forms. The electrolyte flow functions as a slow, dilute "HF rinse" for the cavity. The flow rate used is 3.5 Liters/minute and the rotation speed is 4 revolutions/minute. During this time the electrolyte is warmed to 22 C, resulting in increased draining efficiency when emptying the cavity of electrolyte.

*2.6. Hydrogen uptake*

During the time that the electropolish voltage is applied there is no hydrogen uptake in the niobium. The barrier potential prevents $H^+$ ions from entering the niobium. In order to prevent uptake during the dilute HF rinse phase, a small negative potential is maintained at the cathode. A value of -0.125 V is applied to the cathode from the time prior to introduction of electrolyte into the cavity until the cavity has been emptied of electrolyte and rinsed with water. No instance of Q-disease has been observed in cavities electropolished with this procedure.

*2.7. Voltage*

Table 1 is a selected list of both actively controlled process parameters and the reactive values.

| Parameter | Value |
| --- | --- |
| *Voltage* | 18 V |
| Current | 15 A (average) |
| *Beamtube temperature* | 6 C |
| *Cell temperature* | 16 C |
| *Electrolyte temperature* | 6 C |
| *Electrolyte flow rate* | 1 L/minute |
| *Cavity rotation speed* | 1 revolution/minute |
| *Electrolyte volume* | 12 L |
| Material removal rate | 5 μm/hour |
| Niobium surface current density | 10 mA/cm$^2$ (average) |
| Open cathode area | 100 cm$^2$ |
| Cathode surface current density | 150 mA/cm$^2$ (average) |

Table 1. Elliptical 1-cell cavity electropolish parameters. Parameters under direct control of the operator are shown in *Italics*.

## 3. Results

*3.1. Current waveform*

The current waveform for the set of parameters listed in Table 1 is shown in Fig. 6. The current oscillation is regular and will repeat until the HF in the electrolyte is depleted. For a process removing 30 μm using 12 L of electrolyte, no intermediate adjustment to the parameters will be necessary. For longer procedures, the temperature of the cavity can be increased to restore the oscillation pattern as HF concentration diminishes.

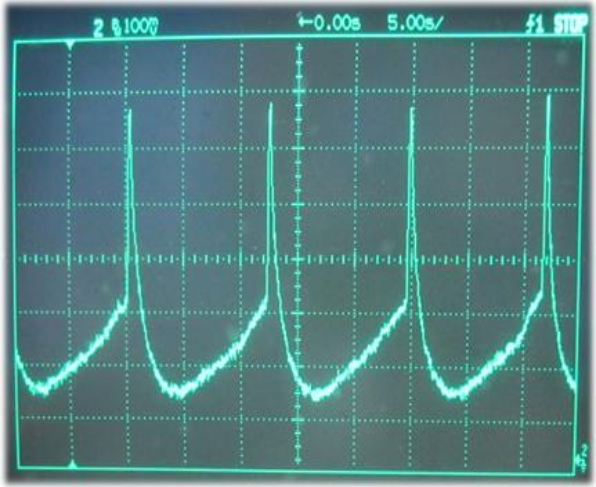

Fig. 6. The current waveform. The scaling for the y-axis is 100 mV = 6 A. The peaks are approximately 40 A. The time period between successive peaks is 12.3 Sec.

It is notable that the depth of the current modulation obtained is 80%. This is considerably more than that shown in Fig 1. The reader should be aware that the y-axis in Fig. 1 does not begin at the value zero. The oscillation depth shown in Fig. 1 has a maximum value of 40%. Both the depth of modulation and the time period for the waveform increase with decreasing cavity wall temperature down to a minimum temperature that is a function of the HF concentration in the electrolyte. At this temperature all oscillation stops. Indeed, there is no current oscillation in the beamtubes during the procedure described here. All the oscillatory behavior occurs in the cavity cell. The indication is that the cell is polishing with a diffusion limiting dielectric layer, as intended.

The concentration of HF in the electrolyte, as received from the vendor, varies within the range of ±30% of the specification. It has, thus far, always been possible to find a suitable set of parameters that result in deeply modulated oscillations. The parameters listed in Table 1 were determined for a batch of electrolyte with HF concentration near the low end of the range of variation. For higher concentrations of HF, the temperatures should be adjusted to lower values. The minimum temperature for the onset of stable oscillations has varied in the range 11 C to 15 C, measured at the cavity equator, between electrolyte batches of different HF concentrations.

## 3.2. Material removal

Material removal is determined by differential measurement of the cavity wall thickness before and after an electropolish procedure. The thickness is determined with an ultrasonic reflectometer that calculates distance from a time-of-flight measurement. With such an instrument a trained and well-practiced operator can achieve an absolute uncertainty of ±5 µm for the thickness of a 3mm specimen of niobium. Throughout this report the term "material removal" is defined as the change in wall thickness near the cell equator. Warning: When comparing removal rates from different laboratories, it is first necessary to determine exactly what is meant by "material removal." Many laboratories use niobium mass lost during an electropolish procedure to define material removal. A value for average lost thickness is then calculated, based on the loss of mass and the interior surface area of the cavity. There can be a factor of two difference in depth of removal according to the definitions. Mass removal measurements tend to overestimate the removal at the equator.

During the electropolish procedure, the current waveform is integrated with respect to time, giving a value for total charge transported in units of Ampere-seconds. A predetermined scale factor converts this value to the material removal at the equator, allowing the process operator to stop the material removal with precision and repeatability on the order of ±0.2 µm. The scale factor is determined for each cavity type and thermal arrangement by means of a single point calibration. Measured material removal is correlated with the integrated value for charge transport. Scale factors are typically determined by performing a cold EP process with 50 µm of removal. The measurement uncertainty of ±5 µm in material removal results in a ±10% uncertainty in material removal calculated using the scale factor. This is the method used for determining material removal for low removal processes. For example, an electropolish process with material removal of 5 µm as calculated from the charge transport scale factor described above will have uncertainty of ±0.5 µm. The scale factor will be significantly different for the case of a 16C procedure and a 30C procedure.

An added benefit of controlled external water cooling is enhanced uniformity of material removal. Fig. 7 shows measured material removal for an electropolish procedure performed at a cell temperature of 21 C. Material removal in the cell varies by ±15% and is significantly lower in the beamtubes. For delicate processes, such as final material removal following nitrogen-doping [5] of a cavity, adequate and uniform material removal is most critical for the high magnetic field region, points 4 through 8. Reaching the correct nitrogen concentration in the RF surface, and therefore the optimum $Q_0$ performance for doped cavities, was achieved at Fermilab by performing sequential material removal and RF test, with removal as low as 1 µm between steps. Reliably small variation in material removal, and therefore nitrogen concentration, was an important asset for these studies.

The rate of material removal for the electrolyte and parameters of Table 1 is approximately 5 µm per hour. This rate is suitable for precision material removal of 10 µm or less, but is too slow for removing approximately 150 µm in the "bulk" removal step for cavities. Fermilab has adopted the procedure of removing the first 140 µm at a temperature of 30 C and then, within the same electropolish process, reducing the temperatures to the values in Table 1. The final 10 µm is then removed with the low temperature settings. The resulting surface is equivalent to a surface that has been polished completely at low temperature. After nitrogen-doping of the cavity, a final -5 µm electropolish is then done using the low temperature technique.

### 3.3. Cavity surface resistance

The cold EP technique was used as the final surface preparation for cavity RDTTD02, a 1-cell 1.3 GHz cavity made by JLab from Tokyo Denkai niobium sheet. The cavity was being used to study the effects of furnace annealing on magnetic flux expulsion during the transition from the normal to the superconducting state at 9.2 K. Prior to its RF test, the cavity was treated as follows:

(1) Electropolish: -30 µm (28 C) -10 µm (16 C)
(2) 900 C (3 hours) + Nitrogen-doping (800 C)
(3) Electropolish: -5 µm (16 C)

$Q_0$ verses $E_{acc}$ performance curves for the cavity are shown in Fig. 8. The maximum value of accelerating gradient ($E_{acc}$) shown in the figure was intentionally limited in order to prevent the chance of a multipacting or nitrogen-doping induced quench, which would increase the surface resistance due to trapped magnetic flux. It is typical for nitrogen-doped cavities to quench in the range of 90 to 110 mT, or, for the TESLA cavity shape used here, in the range of 21 to 25 MV/m

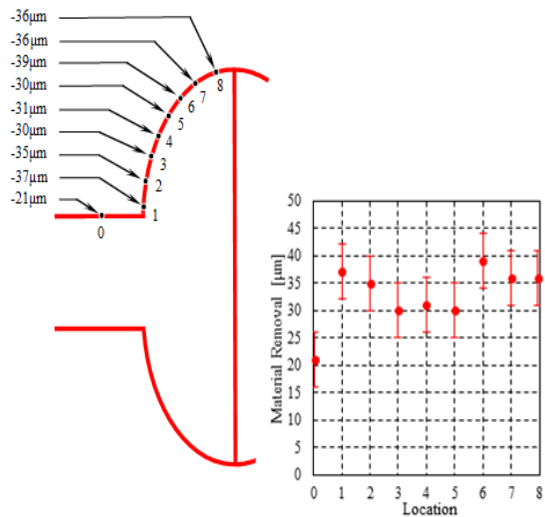

Fig. 7. Elliptical, 1.3 GHz, 1-cell material removal pattern. The cell temperature was 21 C and the beamtube temperature was 8 C for this procedure. Measurement locations 1 through 8 are separated by 1 cm along the outside profile of the cell.

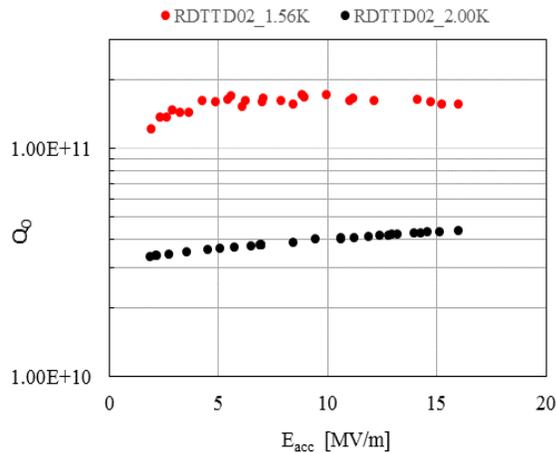

Fig. 8. $Q_0$ verses $E_{acc}$ curves for 2 K and 1.56 K. Uncertainty in the 1.56 K data is approximately ± 30% for $Q_0$ and ±10% for $E_{acc}$.

accelerating gradient.

We would like to determine a reasonable estimate for the residual surface resistance of the cavity based on the RF performance curve. Reference [6] is a good source of information for the physical and mathematical basis for this exercise.

The cavity has a geometry constant equal to 268 Ω. At an accelerating gradient of 10 MV/m, and at 1.56 K the $Q_0$ of the cavity was measured to be $Q_0 = 1.7 \times 10^{11}$. The total surface resistance is therefore $R_s = G/Q_0 = 268\ \Omega\ /1.7 \times 10^{11} = 1.6 \times 10^{-9}\ \Omega$. Approximately $0.6 \times 10^{-9}\ \Omega$ is due to remaining BCS resistance at 1.56 K. (See, for example, Fig 6.2 of [7].) The total residual resistance is then approximately $R_{res} = 1.0 \times 10^{-9}\ \Omega$. After the 900 C treatment, the cavity was measured to be approximately 90% efficient at excluding magnetic flux. For this test, the cavity was cooled through 9.2 K in an average magnetic field of approximately $1 \times 10^{-7}$ T. Therefore, the contribution to surface resistance from trapped flux should be negligible. We can therefore conclude that the contribution to residual surface resistance from the results of the electropolishing procedure is less than or equal to $1 \times 10^{-9}\ \Omega$.

Note that because RDTTD02 is a doped cavity, it did not receive a 120 C x 48 hour low temperature bake. Non-doped cavities that have had a low temperature bake cycle exhibit decreased BCS resistance and increased residual resistance. The additional residual resistance can then be removed by performing an HF rinse on the cavity. Very low residual resistances can be achieved with this technique [8].

Very low added surface resistance is typical of cavities treated with the cold EP method. Equally good results for low surface resistance have, on occasion, been realized from electropolish procedures with substantially different process parameters. Such is the statistical variation of results from differing SRF surface preparation techniques. Here, the claim is made that for a cavity treated according to the cold polish procedure, added surface resistance due to electrochemistry can be reliably managed at the level of $1 \times 10^{-9}\ \Omega$, or less. Cold electropolishing with deeply modulated current can be considered a sufficient condition to achieve low surface resistance, but is not always a necessary condition.

### 4. Extension to multicell cavities

Cooling zone separation can be applied to multicell cavities. The idea is to inject cold water into the holes that are present in intercell strengthening ribs, thus cooling the iris region, while using zone separators on both sides of the rib to prevent the cells from operating at too low a temperature. A lower water flow rate would be used for the cells. A thermocouple would be used to monitor the temperature of each cell. Fig. 9 shows the arrangement of zone separators for a multicell cavity.

It is usual for multicell cavities to be installed in external support frames when they are being electropolished. The presence of such an external structure makes precise control of cooling water problematic. In order to use the technique described in this report an alternative way of supporting the weight of the cavity and the electrolyte that it contains is needed. Fermilab is presently beginning the installation and commissioning of an electropolish tool that allows the center cell of a multicell cavity to be supported by a wheel that rotates with the cavity. A cavity "cage" structure is not used. This allows much closer and more precise approach to the external cavity surface with water nozzles.

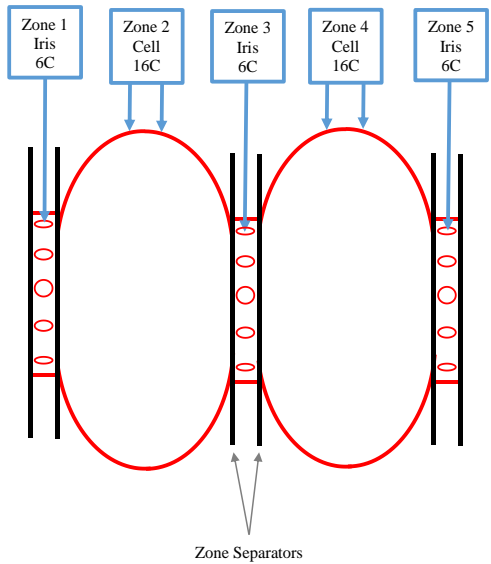

Fig. 9. The cooling zone concept for multicell electropolish.

A significantly different pattern of holes will be required for introducing fresh electrolyte through the cathode. Because electrolyte is usually introduced to the cathode from one side only, the pressure drop due to the dynamic viscosity of the fluid will result in a pressure gradient along the length of the cathode. Holes will need to be sized so that the HF concentration in each cell is approximately equal during the polishing operation.

**5. Open questions**

The current in Fig. 6, at its minimum value, has the same value as for the case of very low electrolyte and cavity temperatures, where no oscillations take place. One interpretation of this is that the niobium surface acts in a unified manner with respect to the HF diffusion limit phenomenon, i.e., there are not separate areas that have their own local oscillating current waveform, with their own distinct phases. Otherwise, the sum of the differing phases, as viewed on the integrated current waveform would not be likely to reach such a low minimum value, nor would the periodicity be uniform. Why such a large surface area would behave in this correlated manner is not known. The fact that large oscillations are present in the "still" electropolishing of Fig. 1 makes any strong influence from cavity rotation speed unlikely.

The logical extension of this question is "How do multicell cavities behave with respect to current oscillation correlation?" Does a cell "know" what its neighbors are doing, and behave with a similar and correlated current waveform? It may be the case that each cell has its own phase for current. This will not result in a clean, easily interpreted integrated waveform. One way to gain information in this situation would be to use the external water cooling system to reduce the temperature of all cells, except one, until only the one cell has current oscillations [9]. For the case of a 9-cell cavity, this would result in the oscillation depth of Fig. 6 added to the minimum baseline current from eight additional cells. This information could then be used to judge the quality of the oscillations. The technique could also be use to electropolish a single cell, or group of cells, should such a need arise.

**6. Conclusion**

Continuous, optimal electropolishing conditions for niobium, as described by Diepers, et al., circa 1971, are regularly achieved in single cell elliptical cavity electropolishing at Fermilab by appropriate adjustment of variable parameters and by application of differential external water cooling. The most critical factors are low concentration of water in the electrolyte, low temperature at the cell surface (12 C to 22 C) and low electrolyte flow velocity. This method can be relied upon to produce an RF surface with a very small ($<1\times10^{-9}\,\Omega$) contribution to surface resistance that is attributable to the material removal procedure.